# FORTNITE & CHEMISTRY

Nicolas DIETRICH

Toulouse Biotechnology Institute (TBI), Université de Toulouse, CNRS, INRA, INSA, Toulouse, France

**ABSTRACT**

In this activity, we describe how the video game *Fortnite* provides a nice opportunity for students to think about chemistry. The game has a survival system based on the acquisition and use of items, which has interesting chemical and physical properties. In addition, the game also provides an important platform of crafting and base building and open discussion on materials.

**KEYWORDS**

General Public, Chemical Engineering, Collaborative / Communication / Writing, Humor / Puzzles / Games, Reactions /History, Philosophy/ Inquiry-Based/Discovery Learning, Physical Properties, Student-Centered Learning

**INTRODUCTION**

The use of high-success movies [1] or trending games [2–6] is a good way get students involved in general chemistry courses. Fortnite is an online video game created in 2017, developed by Epic Games. The game mode includes Fortnite: Save the World, a cooperative shooter-survival game for up to four players, and Fortnite Battle Royale, a free-to-play battle royale game where up to 100 players fight in increasingly-smaller spaces to be the last person standing. The game has cartoon graphics and do no present graphic violence such as blood. While both games have been successful for Epic Games, Fortnite Battle Royale became a resounding success, drawing in more than 125 million players in less than a year, and earning hundreds of millions of dollars per month, and since has been a cultural phenomenon [7].

Fortnite gained enough popularity to actually make students in the classroom excited to play the game and to even care enough about it to carry it over into the classroom. In early 2018, students of Tippecanoe High School in Ohio, USA, used a social media platform as a way of getting out of doing traditional homework about traditional subjects. In fact, one student attempted to get their school to allow them to do a test about Epic Games' sandbox survival game Fortnite for their final exam in chemistry.

The main question of this paper is how to build a chemistry test around Fortnite? The game is more oriented toward physics than chemistry, due to the large amount of properties that could be extract easily centered around physics questions, such as rockets trajectories or the amount of force per impact of certain projectiles, etc. But in the game, after the players has landed on the map, they must scavenge for weapons, resources and items. One of these items is: "*slurp juice*", a consumable that adds shield and health points to the character. This item is represented by a bi-colored viscous fluid with beads in a jar, that could be realized in the chemistry lab with a professor. The fluid can be realized with slime paste with common material (hot water, a spoon of borax, and glue) or chemical product (water, polyvinyl alcohol, and boric acid) in order to illustrate the mechanism of polymerization [8]. Some glass beads and dyes (green for the bottom fluid and blue for the top fluid) can be add before the polymerization in order to improve the resemblance with the real "*slurp juice*" as depicted in figure 1.a.

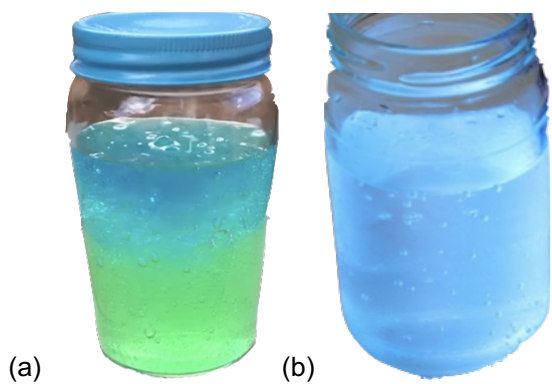

(a)      (b)

Fig 1. Example of Fortnite items that can be realized in the chemistry lab. (a) the "*slurp juice*". (b) the "*shield potion*".

A second famous item in the game is the "*shield potion*", a glowing blue liquid in a jar with gems floating inside. This item can be easily realized using tonic water and a black light. The quinine present in the tonic water will glow with a blue light [9–11], and the carbonic bubbles in the water imitate perfectly the gem present in the shields potion (Figure 1.b). This experiments revealed the phosphorescence properties of quinine but can be done with fluorescein also to enlighten fluorescence effects [12]. This phenomenon can be done also with other products such as energy drinks with B vitamins, milk, vanilla ice cream, caramel, honey to give a yellow color, that could imitate the item "*stink bomb*" in the game, and all these solution can be improved in the chemistry lab with adding few spoon of table vinegar and hydro-gen peroxide or directly with luminol to illustrate the chemiluminescence [13–15]. Another weapon item is the "grenades" that could illustrate the chemistry of the dynamite [16] and its inventor, Albert Nobel [17], the price he founded in 1895 and the synthetic element nobelium was named after him [18]. Finally, as the game also relies on a building mechanic in order to craft fortifications and structures, different material, such as wood, stone, or metal can be granted in destroying some objects already present. This aspect of the game could finally be linked to the chemical structures of these materials [19] or about the chemical engineering needed to produce it [20]. Finally, the game Fortnite offer a wide range of possibility to involved student in chemistry classroom.

**AUTHOR INFORMATION**
Nicolas DIETRICH
*E-mail: nicolas.dietrich@insa-toulouse.fr
Personal website: ndietrich.com
ORCID: orcid.org/0000-0001-6169-3101
Note: The author declares no competing financial interest.